\documentclass{article}
\usepackage{spconf,amsmath,graphicx,hyperref}
\usepackage{cite}
\usepackage{tabularx,booktabs}
\usepackage{multirow}

\usepackage{makecell}

\usepackage{xcolor}

\title{Multiclass Speech Classification Under Noise Disparity}
\name{Mahdi Amiri$^{1 ,2}$, Sayantan Biswas$^{2}$, Mingchi Hou$^{1 ,2}$, Pascal Frossard$^{2}$, Ina Kodrasi$^{1}$ \thanks{This work was supportecd by the Swiss National Science Foundation projects CRSII5\_202228 on ``Characterisation of motor speech disorders and processes'' and 200021\_215187 on ``Pathological Speech Enhancement''.}
}
\address{\textit{$^1$Idiap Research Institute, Switzerland} \\
\textit{$^2$École Polytechnique Fédérale de Lausanne (EPFL), Switzerland}\\}

\begin{document}
\ninept
\maketitle
\begin{abstract}
We investigate noise disparity in multi-class speech classification tasks and develop a strategy to prevent classifiers from exploiting class-dependent noise characteristics. Building on our previous work for binary classification, we introduce a multi-class cross-augmentation scheme that exposes each class to the noise characteristics of the other classes, thereby removing the association between individual noise conditions and class labels. We compare this training-based approach with speech enhancement (SE) as a preprocessing strategy, which aims to suppress noise-related cues directly from the input. Experiments on multi-class emotion recognition show that cross-augmentation effectively mitigates the effect of noise disparity across a range of signal-to-noise-ratios, while SE has a detrimental effect on model performance.


\end{abstract}

\section{Introduction}

Self-supervised learning (SSL) foundation models have emerged as powerful general-purpose speech representations, achieving state-of-the-art performance across a wide range of classification tasks such as emotion recognition~\cite{tilak, ma2024emobox}, pathological speech detection~\cite{amiri2024test, baseline, mel_ref_vae, deep_learning_based}, or speaker verification~\cite{fan2020exploring, kim2025rethinking}. Models such as wav2vec 2.0~\cite{w2v2} and XLS-R~\cite{xlsr} provide rich speech representations without requiring task-specific feature engineering, and have therefore become increasingly prevalent in downstream speech classification tasks.

However, the strong representational capacity of SSL models also makes them susceptible to exploiting \emph{spurious correlations} in the training data. A particularly problematic setting arises when noise characteristics are systematically associated with class labels, a phenomenon we refer to as \emph{noise disparity}. Such disparity can naturally arise when recordings from different classes are collected under different acoustic conditions, with different recording devices or in different environments. In this situation, noise becomes predictive of the class label, allowing a classifier to achieve high performance by exploiting noise-related cues rather than the speech characteristics relevant to the target task. Importantly, this problem can remain hidden when evaluation data share the same noise characteristics as the training data, while causing substantial performance degradation when the acoustic conditions change. 
For example, in~\cite{schu2023using} we showed that binary pathological speech classifiers trained on the TORGO~\cite{torgo} and UASpeech~\cite{uaspeech} datasets perform considerably better on non-speech than on speech segments, demonstrating that they exploit recording-related characteristics present in non-speech segments rather than pathology-relevant properties of the speech signal. 
Similar reliance on non-task-related cues has been observed in other speech classification tasks~\cite{gauder2024unreliability, gauder2026toolkit}.

An intuitive way to mitigate noise disparity is to suppress the noise from the recordings before classification. Speech enhancement (SE) models~\cite{cr_icassp2023, jukic2024schr} are widely used as preprocessing modules for downstream speech tasks~\cite{ SE4ASR_icassp22}, and therefore provide a natural candidate for this purpose. 
In principle, if class-discriminative noise is sufficiently suppressed, the classifier cannot exploit noise-related cues. However, SE does not simply remove noise, as enhancement models can also modify the speech signal and introduce artifacts whose characteristics depend on the input noise, signal-to-noise ratio (SNR), and enhancement model itself~\cite{iwamoto22_artifacts_SE_ASR}. If these artifacts vary systematically with the original noise conditions, they can themselves become predictive of the class labels. Consequently, SE may replace one source of spurious information with another, rather than eliminating the underlying shortcut. 

An alternative strategy is to instead modify the training data itself, removing the association between noise and class label directly.
To suppress noise-related cues for the \emph{binary} pathological speech detection task, we previously proposed an augmentation-based approach that combines each training sample with a proxy of the noise extracted from the opposite class~\cite{amiri_suppress}.
This augmentation exposes both classes to comparable noise characteristics, suppressing the association between noise and class labels so that a classifier can only succeed by relying on genuine speech content.
This formulation, however, is defined for pairs of
samples. In the multiclass setting, a sample can be matched with noise from several other classes, and it is not clear a priori which noise, or which combination of noises, should be added. 


In this paper, we propose an extension of the binary augmentation approach from~\cite{amiri_suppress} to the multiclass setting, using multiclass speech emotion recognition as an exemplary task. Additionally, we investigate whether SE can provide an alternative means of mitigating noise-related shortcuts. Our main contributions are as follows:
\begin{itemize}
\item We demonstrate that noise disparity can induce spurious class-discriminative cues in multiclass speech classification, resulting in models that rely on noise characteristics rather than task-relevant speech information.

\item We extend our augmentation strategy~\cite{amiri_suppress} from binary to multiclass classification by comparing three cross-class noise augmentation schemes. We show that augmenting each sample with noise from all other classes effectively suppresses the noise-disparity shortcut across SNR conditions, while random and batch-pairing strategies are ineffective.

\item We systematically investigate SE as a commonly used preprocessing strategy for handling noisy inputs. We show that in the presence of noise disparity, SE models introduce spurious enhancement-dependent artifacts that also impair classification performance.
\end{itemize}

\section{Preliminaries}
\label{sec: model}

Consider a $C$-class speech classification task, where $y_i^c(k)$ denotes the $i$-th noisy speech sample 
from class $c$ at time index $k$. Each noisy sample is modeled as
\begin{equation}
y_i^c(k) = s_i^c(k) + n_i^c(k),
\end{equation}
where $s_i^c(k)$ and $n_i^c(k)$ denote the corresponding clean speech 
and noise signals, respectively. 
In practice only the noisy mixture $y_i^c(k)$ is observed, while the clean speech and noise components are not available.
We define \emph{noise disparity} as a setting in which the noise distribution depends on the class $c$, i.e., samples belonging to the same class are corrupted according to the same noise distribution, while the noise distribution differs across classes.
This happens when  recordings from different classes are collected under different acoustic conditions. Consequently, noise characteristics can become correlated with the class label and provide spurious cues for classification.

To characterize the signal and noise levels, we define the average power of a signal $x(k)$ of length $K$ as
\begin{equation}
\label{eq_p}
P_x = \frac{1}{K}\sum_{k=0}^{K-1} x^2(k).
\end{equation}
Accordingly, for a noisy signal $y(k)=s(k)+n(k)$, we define the SNR as
\begin{equation}
\label{eq_snr}
\mathrm{SNR}_y^n = 10\log{10}\frac{P_s}{P_n}.
\end{equation}
The notations $P_x$ and $\mathrm{SNR}_y^n$ are used throughout the remainder of this paper to denote the corresponding signal powers and SNRs, respectively. For ease of readability, the time index $k$ is omitted in the following.


\section{Noise disparity in the two-class setting}
\label{sec: two-class}

We proposed a strategy for mitigating noise disparity in the two-class classification setting in~\cite{amiri_suppress}. The underlying idea is to cross-augment samples from the two classes with each other's noise, such that the resulting samples have matched noise characteristics. 
If this is achieved, noise is no longer predictive of the class label and a classifier can then only succeed by relying on genuine speech content.

Specifically, consider a sample $y_i^a$ from class $c_a$ and a sample $y^b_j$ from the other class $c_b$, with $a \neq b$. Their augmented versions are given by
\begin{align}
\label{eq:augment_}
 {\hat{y}^{a}}_{i} = \underbrace{ s^{a}_{i} + n^{a}_{i} }_{y^{a}_{i}} + \Gamma_{b_j \rightarrow a_i} {\hat{n}^{b}}_{j}, \; \;
 \hat{y}^{b}_{j} = \underbrace{s^{b}_{j} + n^{b}_{j}}_{y^{b}_{j}} + \Gamma_{a_i \rightarrow b_j} \hat{n}^{a}_{i},
\end{align}
where $\hat{n}^{a}_{i}$ and $\hat{n}^{b}_{j}$ are the estimated 
noises from samples $y^{a}_{i}$ and $y^{b}_{j}$, respectively, and $\Gamma_{b_j \rightarrow a_i}$ and $\Gamma_{a_i \rightarrow b_j}$ are scalar coefficients used to scale these noise estimates when augmenting each sample. Following~\cite{amiri_suppress}, noise is estimated using the non-speech segments of each sample, which are extracted with a voice activity detector (VAD)~\cite{vad_ref}.
For the two augmented samples ${\hat{y}^{a}}_{i}$ and $\hat{y}^{b}_{j}$ in~(\ref{eq:augment_}) to have the same noise characteristics, the scalars $\Gamma_{b_j \rightarrow a_i}$ and $\Gamma_{a_i \rightarrow b_j}$ should be computed such that (i) the SNR of ${\hat{y}^{a}}_{i}$ and $\hat{y}^{b}_{j}$ with respect to each of the individual noises is the same and (ii) the SNR of ${\hat{y}^{a}}_{i}$ and $\hat{y}^{b}_{j}$ with respect to the total noise is the same.
These conditions are expressed through the system of equations:
\begin{equation}
\left\{
\begin{aligned}
\label{eq: condition1}
\text{SNR}_{\hat{y}^{a}_{i}}^{n^{a}_{i}} & = \text{SNR}_{\hat{y}^{b}_{j}}^{\Gamma_{a_i \rightarrow b_j}{\hat{n}^{a}}_{i}} \\
\text{SNR}_{{\hat{y}^{a}}_{i}}^{\Gamma_{b_j \rightarrow a_i} {\hat{n}^{b}}_{j}} & = \text{SNR}_{{\hat{y}^{b}}_{j}}^{n^{b}_{j}} \\
\text{SNR}_{{\hat{y}^{a}}_{i}}^{n^{a}_{i} + \Gamma_{b_j \rightarrow a_i} {\hat{n}^{b}}_{j}} & = \text{SNR}_{{\hat{y}^{b}}_{j}}^{n^{b}_{j} + \Gamma_{a_i \rightarrow b_j}{\hat{n}^{a}}_{i}}
\end{aligned}
\right.
\end{equation}
As shown in~\cite{amiri_suppress}, the scalars $\Gamma_{a_i \rightarrow b_j}$ and $\Gamma_{b_j \rightarrow a_i}$ satisfying~(\ref{eq: condition1}) are given by
\begin{equation}
\label{eq: alpha_condition}
\Gamma_{a_i \rightarrow b_j} = \sqrt{\frac{{P_{s}}^{b}_{j}}{{P_{s}}^{a}_{i}}}, \; \; \Gamma_{b_j \rightarrow a_i} = \sqrt{\frac{{P_{s}}^{a}_{i}}{{P_{s}}^{b}_{j}}},
\end{equation}
ensuring that the two augmented samples ${\hat{y}^{a}}_{i}$ and $\hat{y}^{b}_{j}$ have matched noise characteristics. Thus, in the two-class setting, each sample can be cross-augmented with a noise estimate from the other class using a uniquely defined scaling coefficient.

\section{Noise disparity in the multiclass setting}
\label{sec: multiclass}
For $C>2$ classes, the formulation of Section~\ref{sec: two-class} does not specify which noise a sample should be augmented with, since the noises of $C-1$ other classes are available. The most direct extension is to augment
each sample with the estimated noise of every other class, so that all augmented samples contain the same $C$ noise sources. Generalizing~(\ref{eq: condition1}), the augmented samples have matched noise
characteristics if their SNR is the same with respect to each individual noise source and with respect to any combination of noise sources, which amounts to $2^C-1$ conditions. Under the same assumptions as in~\cite{amiri_suppress}, these conditions are satisfied by the pairwise coefficients of~(\ref{eq: alpha_condition}). 
However, stacking several noise signals onto a single sample lowers its effective SNR, with the reduction increasing as the number of classes grows. We therefore also consider two variants that randomly introduce noise from only a single class per sample. In this case, the formulation is equivalent to considering only pairwise conditions as in the binary classification setting, without a guarantee of matching all noise characteristics across all classes. In addition, we evaluate SE as an alternative approach that aims to suppress noise entirely prior to classification, rather than balancing it across classes. \\

\noindent 1. \textit{Augmentation with noise from all other classes.} \enspace
Each noisy sample is augmented with a scaled noise estimate from every other class, i.e.,
\begin{align}
\label{eq:augment}
 {\hat{y}^{a}}_{i} = \underbrace{s^{a}_{i} + n^{a}_{i}}_{y^{a}_{i}} + \sum_{b \neq a} \Gamma_{b_{j} \rightarrow a_i} {\hat{n}^{b}}_{j}, 
\end{align}
with each $\Gamma_{b_{j} \rightarrow a_i}$  determined as in~(\ref{eq: alpha_condition}). Since every augmented sample contains contributions from the noise of all classes and satisfies all possible conditions, no individual noise pattern remains distinctive of any particular class.
However, because multiple noise components are added at once, the SNR of the augmented sample decreases as the number of classes increases. \\

\noindent 2. \textit{Random augmentation with a single noise.} \enspace At each training epoch, one noise sample is drawn from a randomly selected different class and added using~\eqref{eq: alpha_condition}. Thus, only one noise component is introduced at a time, and the SNR does not degrade as the number of classes increases. In contrast to the previous strategy, this approach does not satisfy all possible conditions in the multiclass setting and therefore does not guarantee effective suppression of noise disparity. \\


\noindent 3. \textit{Pairing samples in each batch.} 
Within each training batch, samples from different classes are randomly paired and cross-augmented via~\eqref{eq:augment_}, which applies the exact symmetric two-class matching condition rather than the random matching used in the previous strategy.
Samples that cannot be paired within a batch are instead assigned noise from a randomly chosen sample of another class. We consider this variant in addition to strategy 2 to test whether enforcing an exact mutual noise match for at least some class pairs per batch is more effective than the weaker additive exposure in strategy 2. \\

\noindent 4. \textit{Using speech enhancement as a preprocessing step.} \enspace
An alternative approach to mitigating noise disparity is to suppress noise before training. To the best of our knowledge, the effectiveness of SE as a preprocessing step for this purpose has not been systematically studied. We therefore apply an enhancement model to the noisy samples prior to training the model, aiming to reduce class-specific noise cues. Specifically, we consider a complex-valued spectrogram-based regression model (CR)~\cite{cr_icassp2023} and a Schr{\"o}dinger Bridge-based model (SB)~\cite{jukic2024sb}, representing discriminative and generative state-of-the-art SE paradigms. 
Their implementation and training configurations are detailed in Section~\ref{sec:SE_setting}.

\section{Experimental Settings}
\label{sec: exp_setting}
\subsection{Databases}
To investigate noise disparity in a controlled setting, we require a dataset in which the association between class labels and noise characteristics can be explicitly controlled. 
To this end, we follow the procedure of \cite{amiri_suppress} and construct a synthetic noisy dataset by adding selected environmental noises to clean speech samples. This allows us to introduce noise disparity while keeping the underlying speech signal unchanged. 

\emph{Clean dataset.} \enspace
For the clean dataset, we use the IEMOCAP dataset~\cite{busso2008iemocap} which is a widely used benchmark for multiclass emotion recognition. IEMOCAP comprises five sessions, each featuring one male and one female speaker. Although the original dataset contains ten emotion classes, we follow the commonly used four-class protocol, considering the angry, neutral, sad, and happy classes (with excited merged into happy)~\cite{ma2024emobox, tilak}. These four classes comprise $5,531$ samples with a total duration of approximately $7$ hours.

\emph{Noise dataset.} \enspace For generating the noisy samples, we use the DEMAND noise dataset~\cite{thiemann2013diverse}. Specifically, we select the DKITCHEN, NPARK, OMEETING, and PCAFETER noises. Each noise type is assigned to exactly one emotion class, such that the noise characteristics are systematically associated with the class labels.
We consider three scenarios corresponding to high, moderate, and random SNRs:
\begin{itemize}
\item[A.] Samples from all classes are corrupted with their respective noises at an SNR of $20$~dB.
\item[B.] Samples from all classes are corrupted with their respective noises at an SNR of $10$~dB.
\item[C.] Samples from all classes are corrupted with their respective noises at an SNR randomly chosen from the range of $10$ to $20$~dB.
\end{itemize}

\subsection{Speech enhancement  models}
\label{sec:SE_setting}
We consider two state-of-the-art SE models, the discriminative CR model from~\cite{cr_icassp2023} and the generative SB model from~\cite{jukic2024schr}.  
Both SE models follow a U-Net architecture. We took the best checkpoint trained on the widely used SE benchmark dataset WSJ0-CHiME3~\cite{hou2025influencecleanspeech}. This benchmark is composed of clean speech from the English Wall Street Journal dataset~\cite{garofolo_john_s_csr-i_2007_wsj0} and noise from the ChiME3 dataset~\cite{barker_chime3_2015}.

\subsection{Training Protocol}
For conducting our experiments, we use a wav2vec2-based \cite{w2v2} approach. We obtain the feature representation from the raw audio signal using the last layer of the wav2vec 2.0 base model. 
We then pass these representations through two linear layers, which project the representation into 256 and 2 dimensions, respectively. We also use a ReLU nonlinearity between the linear layers. We segment signals into 5-second chunks with 50\% overlap. The final residual segment of each utterance is also retained.

Training is done using the Adam optimizer with a learning rate of $1 \times 10^{-3}$. We also use \textit{ReduceLROnPlateau} with a patience of 5 and a factor of 0.5. Models are trained until a maximum number of $100$ epochs or until the learning rate decreases below $10^{-4}$ of the initial learning rate. The best model is chosen based on the loss value on the validation set.

Training and evaluation is done through the commonly used leave-one-session-out cross-validation setting~\cite{tilak}. At each fold, we use one of the sessions as the test set, and the remaining four sessions are split into a training set and a validation set with a ratio of 0.8 and 0.2, respectively. Performance is evaluated using the Unweighted Accuracy (UA), which is commonly used for evaluating emotion recognition systems with imbalanced emotion classes~\cite{tilak}.

\begin{table*}[!ht]
\caption{UA (\%) of the standard training approach and of different augmentation strategies under oracle and practical setups.}

\label{tab: results}
\begin{center}
\begin{small}
\begin{tabularx}{\linewidth}{l|X|ccccccc}
\toprule

\multirow{3}{*}{\makecell{Dataset \\ SNR Setting}} &\multirow{3}{*}{Test set}  & \multicolumn{7}{c}{Training Strategy} \\ 
&& \multirow{2}{*}{Standard} & \multicolumn{2}{|c}{All Noise}& \multicolumn{2}{|c}{Random Noise}& \multicolumn{2}{|c}{Batch Pairing}\\

&&&\multicolumn{1}{|c}{Oracle} & \multicolumn{1}{|c}{Practical} &  \multicolumn{1}{|c}{Oracle}& \multicolumn{1}{|c}{Practical} &  \multicolumn{1}{|c}{Oracle}& \multicolumn{1}{|c}{Practical} \\

\midrule


\multirow{2}{*}{20 dB} & \multirow{1}{*}{Noisy} 
& \multicolumn{1}{|c}{$ 82.5 $} & \multicolumn{1}{|c}{$ 53.3 $} & \multicolumn{1}{|c}{$ 48.0 $} & \multicolumn{1}{|c}{$ 74.3 $} & \multicolumn{1}{|c}{$ 71.8 $} & \multicolumn{1}{|c}{$ 75.3 $} & \multicolumn{1}{|c}{$ 73.2 $} \\ 

& \multirow{1}{*}{Clean}

& \multicolumn{1}{|c}{$ 44.2 $} & \multicolumn{1}{|c}{$ 52.3 $} & \multicolumn{1}{|c}{$ 47.8 $} & \multicolumn{1}{|c}{$ 47.9 $} & \multicolumn{1}{|c}{$ 48.1 $} & \multicolumn{1}{|c}{$ 46.9$} & \multicolumn{1}{|c}{$ 49.1 $} \\ \hline



\multirow{2}{*}{10 dB} & \multirow{1}{*}{Noisy} 
& \multicolumn{1}{|c}{$ 94.4 $} & \multicolumn{1}{|c}{$ 44.7 $} & \multicolumn{1}{|c}{$ 50.1 $} & \multicolumn{1}{|c}{$ 86.8 $} & \multicolumn{1}{|c}{$ 89.2 $} & \multicolumn{1}{|c}{$ 86.6$} & \multicolumn{1}{|c}{$ 88.8 $} \\ 

& \multirow{1}{*}{Clean}

& \multicolumn{1}{|c}{$ 28.2 $} & \multicolumn{1}{|c}{$ 45.8 $} & \multicolumn{1}{|c}{$ 37.9 $} & \multicolumn{1}{|c}{$ 37.1 $} & \multicolumn{1}{|c}{$ 28.0 $} & \multicolumn{1}{|c}{$ 36.4$} & \multicolumn{1}{|c}{$ 28.3 $} \\ \hline



\multirow{2}{*}{Random} & \multirow{1}{*}{Noisy} 
& \multicolumn{1}{|c}{$ 88.6$} & \multicolumn{1}{|c}{$ 50.3$} & \multicolumn{1}{|c}{$ 48.8$} & \multicolumn{1}{|c}{$81.7 $} & \multicolumn{1}{|c}{$ 80.7$} & \multicolumn{1}{|c}{$81.1 $} & \multicolumn{1}{|c}{$ 81.9 $}\\ 

& \multirow{1}{*}{Clean}

& \multicolumn{1}{|c}{$38.3 $} & \multicolumn{1}{|c}{$  49.6$} & \multicolumn{1}{|c}{$  45.4$} & \multicolumn{1}{|c}{$47.0 $} & \multicolumn{1}{|c}{$ 44.9 $} & \multicolumn{1}{|c}{$46.1 $} & \multicolumn{1}{|c}{$ 44.1 $}\\ \hline



\bottomrule
\end{tabularx}
\end{small}
\end{center}
\vspace{-.7cm}
\end{table*}

\begin{table}[!ht]
\caption{UA (\%) using SE as a preprocessing step prior to training.}
\label{tab: se}
\begin{center}
\begin{small}
\begin{tabularx}{\linewidth}{c|X|cc}
\toprule

\multirow{2}{*}{\makecell{Dataset \\ SNR Setting}} &\multirow{2}{*}{Test set}  & \multicolumn{2}{c}{Enhancement} \\ 
&& \multicolumn{1}{|c}{CR} &  \multicolumn{1}{|c}{SB}\\

\midrule

\multirow{4}{*}{20 dB} & \multirow{1}{*}{Noisy} 
&\multicolumn{1}{|c}{$ 39.5 $} & \multicolumn{1}{|c}{$ 33.8 $}  \\

& \multirow{1}{*}{Clean}
&\multicolumn{1}{|c}{$ 27.7 $} & \multicolumn{1}{|c}{$ 35.6 $}   \\

& \multirow{1}{*}{Enhanced SB} 
&\multicolumn{1}{|c}{$ 64.1 $} & \multicolumn{1}{|c}{$ 76.1 $}   \\

& \multirow{1}{*}{Enhanced CR} 
&\multicolumn{1}{|c}{$ 75.4 $} & \multicolumn{1}{|c}{$ 61.8 $}   \\ \hline

\multirow{4}{*}{10 dB} & \multirow{1}{*}{Noisy} 
&\multicolumn{1}{|c}{$ 52.7 $} & \multicolumn{1}{|c}{$ 31.9 $}  \\

& \multirow{1}{*}{Clean}
&\multicolumn{1}{|c}{$ 31.1 $} & \multicolumn{1}{|c}{$ 31.5 $}   \\ 

& \multirow{1}{*}{Enhanced SB} 
&\multicolumn{1}{|c}{$ 61.2 $} & \multicolumn{1}{|c}{$ 79.4 $}   \\

& \multirow{1}{*}{Enhanced CR} 
&\multicolumn{1}{|c}{$ 85.4 $} & \multicolumn{1}{|c}{$ 74.6 $}   \\ \hline

\multirow{4}{*}{Random} & \multirow{1}{*}{Noisy} 
&\multicolumn{1}{|c}{$ 44.9$} & \multicolumn{1}{|c}{$ 36.0$}  \\

& \multirow{1}{*}{Clean}

&\multicolumn{1}{|c}{$  27.8$} & \multicolumn{1}{|c}{$ 35.7$}   \\ 

& \multirow{1}{*}{Enhanced SB} 
&\multicolumn{1}{|c}{$ 64.4$} & \multicolumn{1}{|c}{$ 76.3$}   \\

& \multirow{1}{*}{Enhanced CR} 
&\multicolumn{1}{|c}{$ 79.8$} & \multicolumn{1}{|c}{$ 69.6$}   \\ \hline

\bottomrule
\end{tabularx}
\end{small}
\end{center}
\vspace{-.7cm}
\end{table}

\section{Experimental Results}

In this section, we compare the effectiveness of different training strategies for mitigating the effects of noise disparity. As a baseline, we consider standard training in which the model is trained directly on the noisy training data without any augmentation or preprocessing. This baseline is compared with the proposed augmentation strategies.
In addition, we investigate the SE-based training strategy where the noisy training data are first enhanced using either the CR or SB model and the classifier is subsequently trained on the enhanced data. Furthermore, we train and test the model on clean data to obtain a sense of its performance when there is no noise disparity in the dataset.  \emph{The obtained UA in this case is $\mathbf{59.3\%}$}.

For the augmentation-based approaches described in Section~\ref{sec: multiclass}, augmenting each signal with noise from another class according to~(\ref{eq:augment}) requires an estimate of the corresponding noise signal. To investigate the effect of potential errors in the estimated noise signals, we consider two settings:
\begin{enumerate}
    \item \textit{Oracle.} \enspace Since we generate the synthetic noisy dataset ourselves, the exact noise signal in each sample is known. In the oracle setting, we use this ground-truth noise signal during augmentation. This setting allows us to evaluate the augmentation strategy without errors in the noise estimate.
    \item \textit{Practical.} \enspace In practice, the noise signal is unknown and must be estimated from the noisy speech. Following \cite{amiri_suppress}, we use the VAD 
    from \cite{vad_ref} to separate speech and non-speech segments of the noisy samples. Assuming that the non-speech segments contain no speech information, we use them as an estimate of the background noise for augmentation. 
Errors in this estimate therefore reflect the noise estimation errors encountered in a practical setting.
\end{enumerate}

All models are evaluated on two common test sets, i.e., the original noisy test set, which exhibits the same class-dependent noise as the training data, and the corresponding clean test set. A model that relies only on speech content cannot have a higher accuracy on the noisy test set than on the clean one. A higher accuracy on the noisy test set therefore indicates reliance on class-dependent noise cues, and the gap between the two serves as our main measure of spurious-cue exploitation. Since SE-based models are trained on enhanced data, and would also receive enhanced inputs at deployment, we additionally evaluate them on the noisy
test set enhanced with the SB and the CR model. The clean-data reference (UA of $59.3\%$) indicates what can be achieved from speech content alone. Accuracy on enhanced data exceeding it therefore indicates that the classifier exploits cues introduced or preserved by SE rather than
task-related information. Results for the augmentation strategies are given in Table~\ref{tab: results}, and for the SE-based strategies in Table~\ref{tab: se}.


As can be seen in Table \ref{tab: results}, when the model is trained under standard training, its performance on the noisy test set is much higher than on the clean test set across all SNR settings. This confirms that the model learns noise-discriminant cues rather than actual task-related cues.
When augmenting each sample with oracle noise from every other class, our proposed method can effectively close the performance gap between the noisy and clean test sets. This shows that given a precise estimate of the noise, our proposed method can make training robust to spurious correlations. Under the practical setup in the SNR setting of $20$ dB, we also successfully close the gap between the noisy and clean test sets, though this comes at the cost of lower overall performance compared to the oracle setup. Under the practical setup in the SNR settings of $10$ dB and random, the gap between noisy and clean test set performance is reduced by a large margin but does not fully close. We argue that with a better estimate of the noise, this remaining gap could be closed to match that of the oracle scenario.

In the training strategy in which noise from another class is randomly selected for each sample, clean set accuracy increases and noisy set accuracy decreases compared with standard training. However, this strategy is overall not effective, with the performance gap between noisy and clean test sets remaining large. This trend can also be observed in the strategy where we pair samples within each batch and cross-augment their noise. We argue this occurs because the noise is changed for each sample at every training epoch, encouraging the model to become invariant to the augmented noise and instead focus on the components that remain fixed across epochs. As a result, the model can still exploit the original class-dependent noise characteristics, leading to a persistent performance gap between the noisy and clean test sets.

The enhancement-based strategies in Table \ref{tab: se} show a different behavior. While training with either CR or SB results in low performance on the noisy and clean test sets, performance is substantially higher on the enhanced test sets, particularly when the same enhancement model is used for training and testing. Notably, performance on the enhanced test sets is also substantially higher than when training and testing on the clean data (i.e., $59.3$\%). Since the clean data contain only the task-relevant speech information in our controlled setup, the additional discriminative information in the enhanced data must originate from the enhancement of the noisy signal. Thus, rather than removing the class-dependent noise cues, SE appears to preserve or introduce noise-dependent artifact cues that can be exploited by the classifier. Consequently, SE does not provide an effective solution for suppressing noise disparity.

In summary, experimental results show that augmenting each sample with the noise of every other class is the
only strategy that removes the reliance on class-dependent noise cues in multiclass settings. In contrast, single-class augmentation leaves a substantial gap between noisy and clean test set performance, while SE introduces new spurious cues
rather than removing existing ones.

\section{Conclusion}

We investigated noise disparity between classes in multiclass speech classification, using a synthetic dataset built by introducing controlled noise disparity into the IEMOCAP corpus~\cite{busso2008iemocap}. Evaluating this setup across several SNR conditions and test sets, we showed that noise disparity leads models to learn spurious, noise-based cues rather than task-relevant information. To address this, we proposed and compared three cross-augmentation strategies, finding that augmenting each sample with noise from every other class provides the strongest robustness. We further showed that speech enhancement applied as a preprocessing step does not resolve this issue, but instead introduces its own noise- and enhancement-specific artifacts.


\bibliographystyle{IEEEbib}
\bibliography{refs}


\end{document}